\documentclass[aps,prl,twocolumn,groupedaddress]{revtex4}
\usepackage{epsfig,amsmath,amssymb,amsfonts,bm,subfigure,setspace}

\begin{document}
\newcommand{\be}{\begin{equation}}
\newcommand{\ee}{\end{equation}}
\newcommand{\bea}{\begin{eqnarray}}
\newcommand{\eea}{\end{eqnarray}}
\newcommand{\nn}{\nonumber}
\newcommand{\ba}{\bea \begin{array}}
\newcommand{\ea}{\end{array} \eea}
\renewcommand{\(}{\left(}

\renewcommand{\)}{\right)}
\renewcommand{\[}{\left[}
\renewcommand{\]}{\right]}
\newcommand{\bc}{\begin{center}}
\newcommand{\ec}{\end{center}}
\newcommand{\p}{\partial}

\newcommand{\red}{\textcolor{red}}
\newcommand{\mb}[1]{ \mbox{\boldmath$#1$}}
\newcommand{\ds}{\displaystyle}
\newcommand{\beq}{\begin{eqnarray}}
\newcommand{\eeq}{\end{eqnarray}}
\newcommand{\beqq}{\begin{eqnarray*}}
\newcommand{\eeqq}{\end{eqnarray*}}

\newcommand{\g}{\gamma}
\newcommand{\epsv}{\epsilon}
\newcommand{\eps}{\varepsilon}

\newcommand{\x}{\mbox{\boldmath$x$}}
\newcommand{\hx}{\mbox{\boldmath$\hat x$}}
\newcommand{\n}{\mbox{\boldmath$n$}}
\newcommand{\J}{\mbox{\boldmath$J$}}
\newcommand{\Sb}{\mbox{\boldmath$S$}}
\newcommand{\y}{\mbox{\boldmath$y$}}
\newcommand{\z}{\mbox{\boldmath$z$}}

\title{Transcription factor search for a DNA promoter in a three-states model}

\author{J\"urgen Reingruber and David Holcman}
\affiliation{Department of Computational Biology, Ecole Normale Sup\'erieure, 46 rue
d'Ulm, 75005 Paris, France.}

\begin{abstract}
To ensure fast gene activation, Transcription Factors (TF) use a mechanism known as
facilitated diffusion to find their DNA promoter site. Here we analyze such a process
where a TF alternates between 3D and 1D diffusion. In the latter (TF bound to the DNA), the TF further switches between a fast translocation state dominated by interaction with the DNA backbone, and a slow
examination state where interaction with DNA base pairs is predominant. We derive a new formula for the mean search time, and show that it is faster and less sensitive to the binding energy fluctuations compared to the case of a single sliding state. We find that for an optimal search, the time spent bound to the DNA is larger compared to the 3D time in the nucleus, in agreement with recent experimental data. Our results further suggest that modifying switching via phosphorylation or methylation of the TF or the DNA can efficiently regulate transcription.
\end{abstract}

\maketitle


Transcription factors (TFs) are messengers regulating
gene activation by binding the DNA at specific promoter
sites. Interestingly, both theoretical and experimental evidences show
\cite{HippelBerg_JBC1989,Halford_Review_NuclARes2004,WangAustinCox_PRL2006,
BlaineyXie_Sliding_PNAS2006,Elf_Science2007}
that a TF finds rapidly its promoter site by facilitated diffusion,
where it alternates between a 3D diffusion inside the nucleus and a
1D diffusion (sliding) along the DNA strand. Facilitated diffusion
was introduced to resolve the apparent paradox that the measured
in-vitro association rate of the Lac-I repressor with its promoter
site placed on $\lambda$-phage DNA \cite{Riggs_JMolBiol1970} was
$k_R \sim 10^{10}\, (M\,s)^{-1}$, which is $\sim 100$ times larger
than the Smoluchowski rate for a pure 3D diffusion search. However,
the in-vivo mean time $\tau$ for the Lac repressor to find its
promoter site in E-Coli is around $350s$
\cite{Elf_Science2007}, from which we estimate that the association
rate in a nucleus with volume $|V|\sim 1\mu m^3$ is approximated by
$k_{E}=N_{Av}|V|/\tau \sim 10^6 (M\,s)^{-1}$ ($N_{Av}$ is the Avogadro constant). The difference
$k_E \ll k_R$ is due to a slow 1D motion
\cite{Elf_Science2007,WangAustinCox_PRL2006}, such that frequent
non-specific bindings with the DNA in a crowded nucleus slow down the search  and reduce the
association rate. Theoretical analysis
\cite{SlutskyMirny_BJ2004,ZwanzigPNAS1988} shows that the effective
1D diffusion constant for sliding along the DNA decays exponentially
with the variance $\sigma$ of the binding energy distribution
between a TF and the underlying DNA, and a realistic search time can
only be achieved for smooth profiles with $\sigma \lesssim 1.5k_BT$
\cite{SlutskyMirny_BJ2004}. However, binding energy estimations for
the {\it Cro} and {\it PurR} TF on E. Coli DNA
\cite{SlutskyMirny_BJ2004,Gerlandetal_TFinteraction_PNAS2002} show a
much larger variance, suggesting that a simple sliding process is
not sufficient to explain the search dynamics when the TF is bound
to the DNA. In a more complex model
\cite{WinterBergvHippel_3_Biochem1981,SlutskyMirny_BJ2004},
supported by experimental observations \cite{Kalodimos_Science2004},
a TF switches between two conformations when bound to the DNA: in
one state it is insensitive to the underlying DNA sequence and
diffuses quickly in a smooth energy landscape, while in a second
state it interacts with the DNA, reducing the motion. The impact of
such switching has been investigated in
\cite{HuGrosbergBruinsma_BJ2008} based on equilibrium considerations.
In general, switching processes are important because they modulate the rate of chemical
reactions and lead to interesting behavior \cite{Doering_LectureNotes_2000,DoeringGadooua_PRL1992,ReingruberHolcman_PRL2009}.

\begin{figure}[t]
\begin{center}
      \includegraphics[scale=0.20]{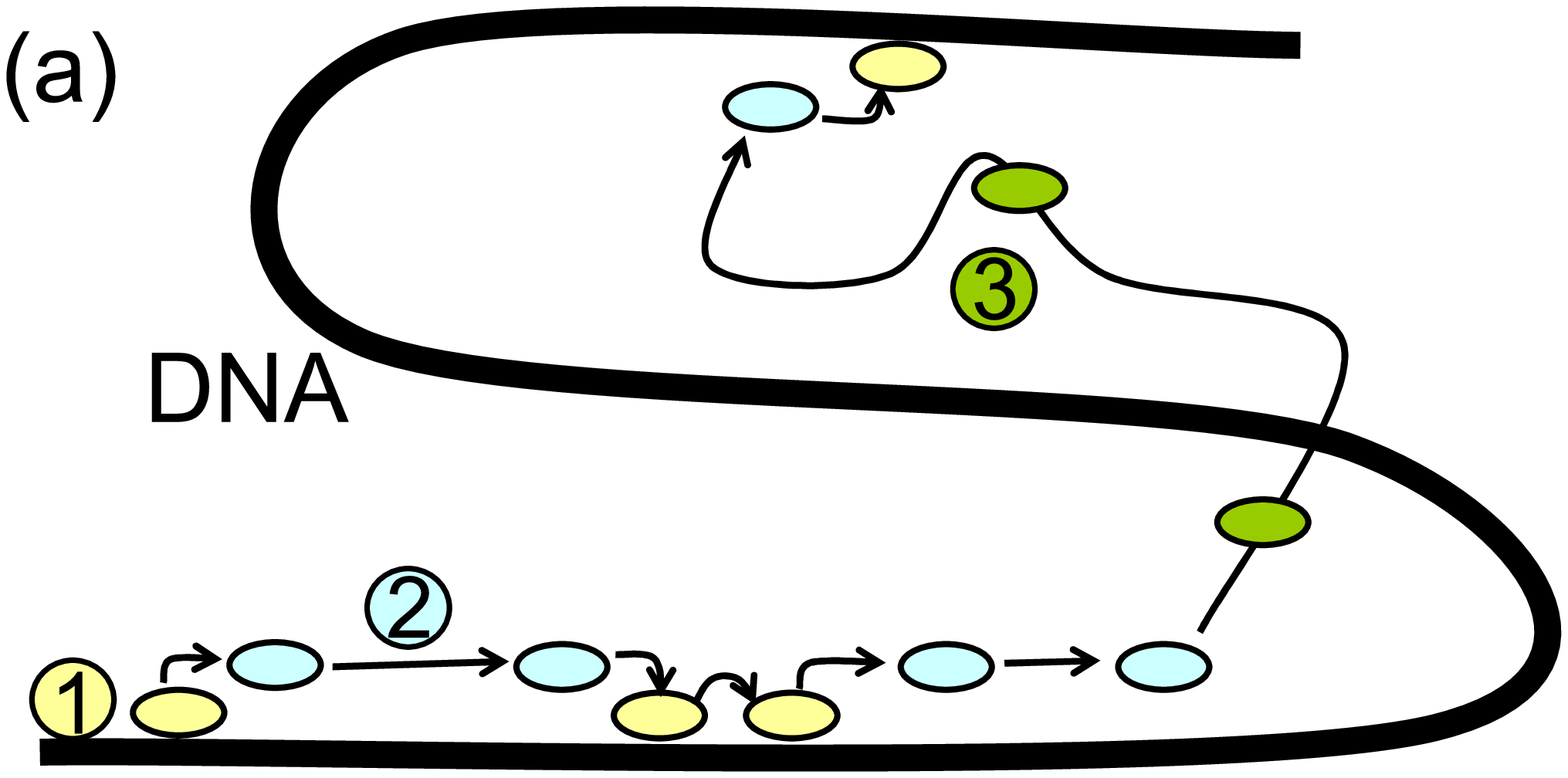}
       \includegraphics[scale=0.20]{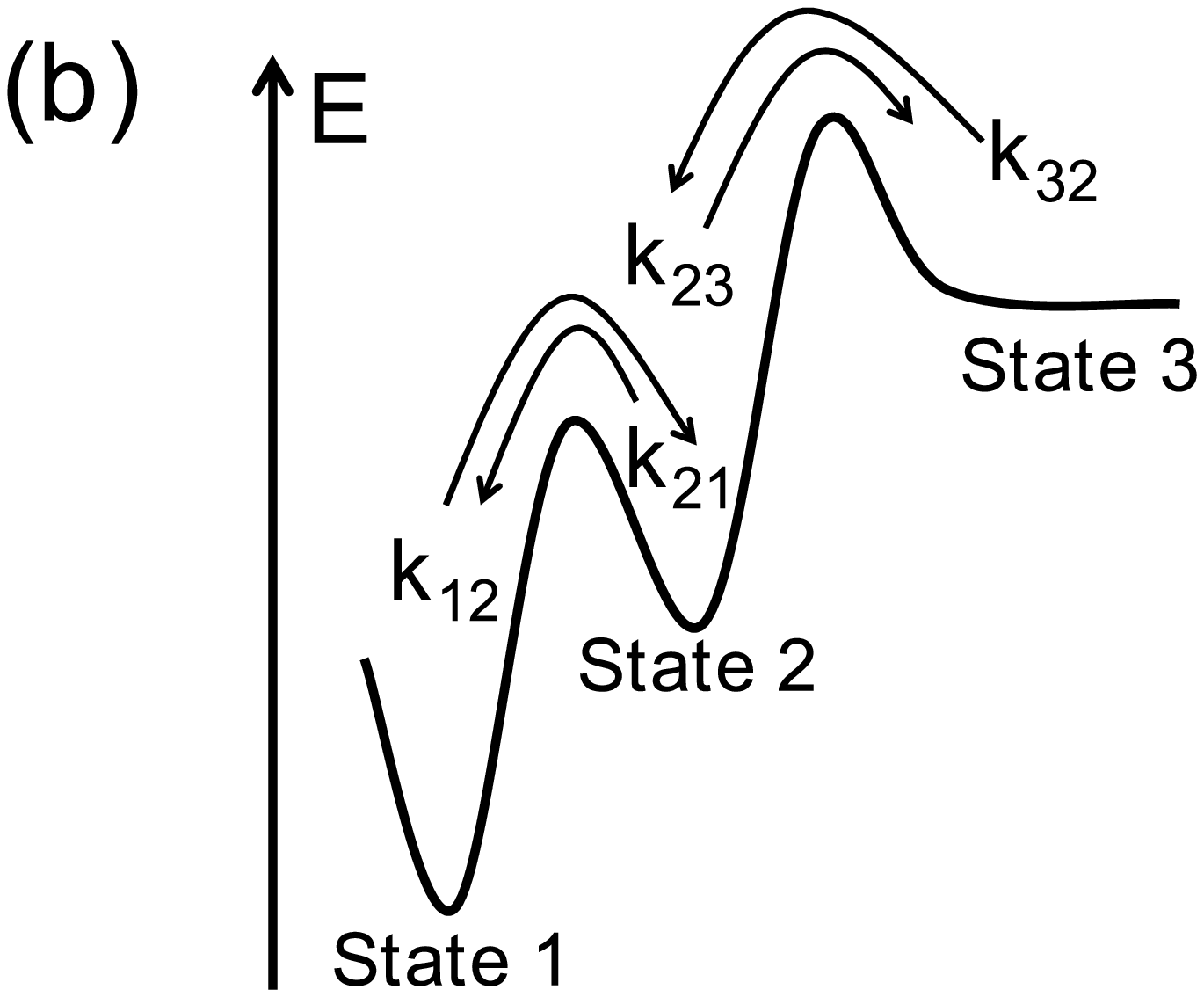}
       \caption{(a) Search scenario with 3 states. (b)
       Energy profile and switching rates between states.}
       \label{fig_scenario}
\vspace{-0.7cm}
\end{center}
\end{figure}

Here we study the mean first passage time (MFPT) for a TF to bind to
its promoter site when it freely moves in the nucleus, but once
bound to the DNA, it alternates between two states (Fig.~\ref{fig_scenario}): in state 1, it interacts with
individual bp, while in state 2 it is insensitive to the underlying
bp sequence and interacts with the DNA backbone. Therefore, in state 1 motion occurs in a rough energy landscape
approximated by an effective diffusion with a slow diffusion
constant $D_1$, while in state 2 diffusion is faster ($D_2 \gg
D_1$) and occurs in a smooth potential well generated by the
interaction with the DNA backbone. The translocations in state 2 are comparable to 'hoppings' along the
DNA. The switching dynamics is Poissonian with rates $k_{12}$ and $k_{21}$ that depend on the
energy profile (Fig.~\ref{fig_scenario}b). In general, the binding
time $k_{12}^{-1}$ depends on the DNA sequence and therefore on the
position along the DNA, however, in first approximation, we use
a constant value. In state 2, in addition to switching to state 1,
the TF can detach from the DNA with rate $k_{23}$ and switch to
state 3, where it diffuses in the nucleus before reattaching
in state 2 after an average time $k_{32}^{-1}$, investigated in
\cite{MalherbeHolcman_PLA2010,LomholtBroeckMetzler_PNAS2009,Halford_Review_NuclARes2004,BergEhrenberg_BiophysChem1982}.
Due to the packed and coiled DNA conformation, we approximate the TF
reattachment locations as uncorrelated and randomly distributed
along the DNA \cite{SlutskyMirny_BJ2004,LomholtMetzler_PRL2005,CoppeyBenichou_TFSearch_BJ2005,Benichou_Traps_PRL2009}. We derive a new expression for the MFPT to find a
promoter site (eq.~\ref{tau_3states}), and we show that
1) this time is not very sensitive to binding energy fluctuations,
contrary to previous models with a single sliding state, and 2) an optimal search process (eq.~\ref{ratio_r_3states}) proceeds such that a TF spends more time bound the DNA compared to freely diffusing in the nucleus, in agreement with recent experiments \cite{Elf_Science2007}.

We start the analysis by considering diffusion along the DNA in the 1D
interval $0 \le x \le L$ ($x$ is the DNA contour length) with switching between state 1 and 2. The target is located at $x=0$ and can only be found in state 1. To derive an expression for the MFPT, we use the sojourn times $t_{nm}(x)$ a particle spends in state $n$ ($n=1,2,3$) when it started in state $m=1,2$ at a DNA position $x$. Because a TF attaches to the DNA at a random position $x$, when starting the search in state 3, the sojourn times do not depend on the initial position, and we have $t_{n3}=\tau_{n3}=const$. The times $\tau_{n3}$ are related to the spatially averaged sojourn times $\tau_{nm}=L^{-1} \int_0^L t_{nm}(x)dx$. Considering that a TF can only bind to the target in state 1, we have the relations $\tau_{13}=\tau_{12}$, $\tau_{23}=k_{12}/k_{21}\tau_{13}+1/k_{21}$ and $\tau_{33}=k_{23}/k_{32}\tau_{23}+1/k_{32}$. The coupled system of equations describing $t_{11}(x)$ and $t_{12}(x)$ is \cite{ReingruberHolcman_JCondM2010} (we suppress the $x$ dependency)
\beq \label{eqsoj3states}
\begin{array}{lll}
D_1 t_{11}'' - k_{12} ( t_{11} - t_{12})  &=& -1  \\
D_2 t_{12}'' - k_{21} ( t_{12}- t_{11})  - k_{23} ( t_{12}- \tau_{12}) &=& 0
\end{array}
\eeq
with boundary conditions  $t_{11}(0)= t_{11}'(L) =
t_{12}'(0)= t_{12}'(L) = 0$. The remaining
sojourn times $t_{2m}(x)$ and $t_{3m}(x)$ are $t_{2m}(x)={k_{12}}/{k_{21}}t_{1m}(x) +
k_{21}^{-1}(1-\delta_{m1})$ and
$t_{3m}(x)={k_{23}}/{k_{32}}t_{2m}(x)$. By integrating eq.~\ref{eqsoj3states} we further obtain the intuitive relation $\tau_{11} = \tau_{12}$. Hence, starting initially uniformly distributed in state $m$, the MFPT
$\tau(m)=\tau_{1m}+\tau_{2m}+\tau_{3m}$ can be expressed in terms of
$\tau_{11}$ only. In particular, starting in state 1, we have $\tau(1) = \tau_{11}
(1+ {k_{12}}/{k_{21}} + {k_{12}}{k_{23}}/({k_{21}}{k_{32}}))$.

Using the variables $\hat
x={x}/{L}$, $l_{12}={k_{12}/(L^2}{D_1})$,
$l_{21}={k_{21}/(L^2}{D_2})$ and $l_{23}={k_{23}/(L^2}{D_2})$, and
the functions $v_1(\hat x) =  k_{12}\tau_{11}(x)$ and $v_2(\hat x) = k_{12}\tau_{12}(x)$
($v_1$ is the mean number of switchings between state 1 and
2), the solutions of eq.~\ref{eqsoj3states} are
\bea\label{solutionv3states}
 \begin{array}{c}
\begin{pmatrix}
\ds v_1(\hat x) \\ v_2(\hat x)
\end{pmatrix}
= \ds \frac{l_{21}}{\xi_2} \( \frac{\cosh(\sqrt{l_{12}} \mu_2 (1-\hat x))}{\sqrt{l_{12}}\mu_2\sinh (\sqrt{l_{12}}\mu_2)} -\frac{1}{l_{12}\mu_2^2}\) \vec e_2 \vspace{0.2cm} \\
\ds - \frac{l_{21}}{\xi_2} \(  \frac{\cosh(\sqrt{l_{12}}\mu_1 (1-\hat x))}{\sqrt{l_{12}}\mu_1\sinh (\sqrt{l_{12}}\mu_1)}-\frac{1}{l_{12}\mu_1^2}\) \vec e_1 + v_1 \,,\nn
\end{array}
\eea
where $\xi_2 = \sqrt{\(1 +({l_{21}+l_{23}})/{l_{12}}\)^2 -4 {l_{23}}/{l_{12}}}$, $\xi_1 =-1 + ({l_{21}+l_{23}})/{l_{12}}$,
$\mu_1^2 = 1+({\xi_1 -\xi_2})/{2}$, $\mu_2^2 = 1+({\xi_1 +\xi_2})/{2}$ and $\vec e_1^{\top} = ({l_{12}}({\xi_1 + \xi_2})/({2} {l_{21}}),1)$, $\vec e_2^\top = ({l_{12}} ({\xi_1 - \xi_2})/({2}{l_{21}}),1 )$. The average $v_1=\int_0^1 v_1(\hat x)d\hat x$ is
\bea\label{v13states}
 \begin{array}{lcr}
\ds v_1 &=&  \ds \frac{\xi_2 -\xi_1}{2\xi_2} \(\sqrt{l_{12}}\frac{\coth(\sqrt{l_{12}} \mu_2)}{\mu_2} -\frac{1}{\mu_2^2} \)  \vspace{0.1cm} \\
&&\ds + \frac{\xi_1 + \xi_2}{2\xi_2} \( \sqrt{l_{12}} \frac{\coth(\sqrt{l_{12}} \mu_1)}{\mu_1} - \frac{1}{\mu_1^2}\)\,. \end{array}
\eea

Because $\xi_1$, $\xi_2$, $\mu_1$ and $\mu_2$ are all independent of
$L$, $v_1$ depends on $L$ only via $l_{12}$. The relevant physical
parameters are $L$, $k_{12}$, $k_{21}$, $k_{23}$, $k_{32}$, $D_1$
and $D_2$. However, to facilitate our further discussion, we shall
now characterize the rates $k_{12}$, $k_{21}$ and $k_{23}$ by the detaching
probability $q={k_{23}}/({k_{21} + k_{23}})$ to switch from state 2
to 3 ($p=1-q$ is the probability to switch from state 2
to 1) and the lengths $l_{s1}=\sqrt{{D_1}/{k_{12}}}$ and
$l_{s2}=\sqrt{{D_2}/({k_{21} + k_{23}})}$, corresponding to the
average sliding distances in state 1 and 2 before switching. The
spatially averaged search time $\tau
\approx \tau(1)$  is
\bea\label{tau_3states}
\tau =  v_1 \(\frac{l_{s1}^2}{D_1} + \frac{l_{s2}^2}{pD_2} + \frac{1}{k_{32}} \frac{q}{p} \) \,.
\eea
Before detaching and switching to state 3, a TF stays bound to the
DNA for an average time $\tau_{DNA}=k_{23}^{-1} + k_{12}^{-1} p/q$,
and the overall ratio of the mean time bound to the DNA to the mean
time spent in state 3 is
\bea\label{ratio_r_3states}
r= k_{32} \tau_{DNA} =  \frac{ k_{32} l_{s1}^2}{D_2} \( \frac{p}{q}  \frac{D_2}{D_1} + \frac{l_{s2}^2}{q l_{s1}^2} \)  \,.
\eea
When switching between state 1 and 2 is fast and diffusion in state
1 is negligible compared to state 2 ($D_1 \ll
D_2$), then the diffusion constant with which a TF appears to slide
along the DNA is
\bea
D_{a} \approx \frac{D_2}{1+ {k_{21}}/{k_{12}}} = \frac{D_2}{1+ p D_2l_{s1}^2/(D_1l_{s2}^2)} \,.
\eea

When the parameters $L$, $D_2$ and $k_{32}$ are given, we shall now
study how the search process depends on $l_{s1}$, $l_{s2}$, $q$
and $D_1$. Modulating these parameters can be a way to regulate gene
expression. Because a TF moves in state 2 in a smooth potential, we
consider that $D_2$ is comparable the 3D diffusion constant. In
contrast, in state 1, the TF interacts with individual bp and the
effective diffusion constant is much reduced and can be written as
$D_1=D_2e^{-\chi}$, where $\chi>0$ depends on the binding energy
profile. For a single sliding state, $\chi$ is related to the
variance of the binding energy
\cite{SlutskyMirny_BJ2004,MalherbeHolcman_PLA2010}. { In general,
$\chi$ depends on the DNA sequences and therefore on the position
along the DNA, however, we consider a constant average value here.
We later on show that the search is not much sensitive to $\chi$
variations in x, as long as $\chi$ is not too large.}  We now
proceed with the asymptotic analysis in the regime where
$\kappa={l_{s1}^2}/{l_{s2}^2}\ll 1$ and $q\ll1$. The condition
$\kappa \ll 1$ avoids a redundant search in state 1 where diffusion
is slow. As long as switching between state 1 and 2 is fast compared
to the time $k_{32}^{-1}$ spent in state 3, the limit $q\ll1$ avoids
too frequent detaching from the DNA that would increase the search
time. Under the condition that $\kappa
\ll 1$ and $ q\ll1$,  we have the asymptotic  $\xi_1 \approx-\( 1 -
\kappa\)$, $\xi_2 \approx 1 +
\kappa( 1 - 2q )$, $\mu_1^2
\approx \kappa q$, $\mu_2^2 \approx 1+ \kappa$ and $v_1 \approx
{L}/{l_{s1}} ( 1 + \sqrt{{\kappa}/{q}})$. Using these expressions
in eq.~\ref{tau_3states} and eq.~\ref{ratio_r_3states}, we find
\bea\label{tau3states_asymp}
\tau &\approx&  \sqrt{\frac{L^2}{D_2k_{32}}} \( 1 + \sqrt{\frac{\kappa }{q}}\)  \(\frac{e^{\chi}}{\alpha}  + \frac{1}{\alpha \kappa} + \alpha q \)\,,
\eea
\bea
r &\approx&  \frac{e^{\chi}}{ \alpha^2 q} +\frac{1}{\alpha^2 q \kappa} \,,
\eea
where $\alpha = \sqrt{{D_2}/({l_{s1}^2k_{32}})}$. When  $\alpha$ and
$\chi$ are fixed, the minimum of $\tau(q,\kappa)$ as a function of
$\kappa$ and $q$ is achieved for
$(\kappa_{min},q_{min})=(\sqrt{{2}/({\alpha e^{\chi}})}, \alpha^{-2}
\kappa_{min}^{-1})$, and
\bea\label{tauminandrmin}
\ds \tau_{min}&=& \ds \sqrt{\frac{L^2}{D_2k_{32}}} \( 1+\sqrt{\frac{2\alpha}{e^{\chi}}} \)^2
\frac{e^{\chi}}{\alpha}\,. \\
r_{min}&=& 1+ \sqrt{{2e^{\chi}}/{\alpha}}.
\eea
{For ${e^{\chi}}/{\alpha} \ll1$, the asymptotic expansion is
$\tau_{min}\approx 2 \sqrt{{L^2}/({D_2k_{32}})}(1+
\sqrt{{2e^{\chi}}/{\alpha}})$, showing that $\tau_{min}$
does not depend exponentially on $\chi$ in that regime.} We now
compare our results with the ones for a single sliding state:  when
a TF alternates only between state 1 and 3 with rates $k_{13}$ and
$k_{31}$ (the intermediate state 2 is absent), we find from
eq.~\ref{v13states} that $\tilde v_1=\sqrt{l_{13}}=
\sqrt{{L^2}/({D_1 k_{13}})}$, and for the search time we recover the
expression $\tilde \tau = \sqrt{{L^2}/({D_1 k_{13}})}
(k_{13}^{-1} + k_{31}^{-1} )$
\cite{MalherbeHolcman_PLA2010,Elf_Science2007,SlutskyMirny_BJ2004,Klafter_JPhysA2009}.
When $k_{31}$ is fixed, the minimum $\tilde
\tau_{min} = 2\sqrt{{L^2}/({D_1 k_{31}})}$ is achieved for
$k_{13}=k_{31}$, and $r$ is always one at the
minimum with a single sliding state, which is not any longer the case in the two states sliding model.

\begin{figure}[h!]
\begin{center}
      \includegraphics[scale=0.28]{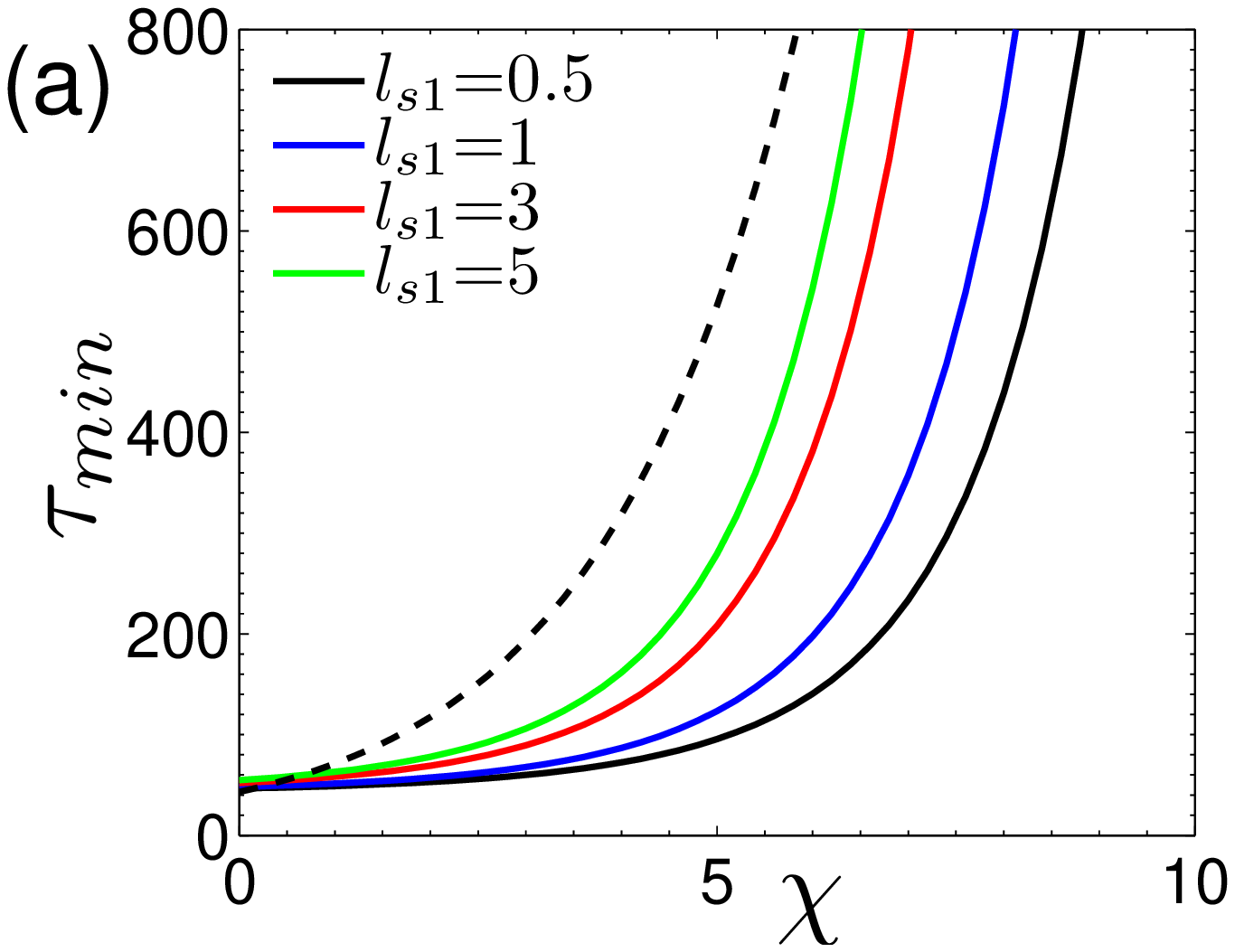}
       \includegraphics[scale=0.28]{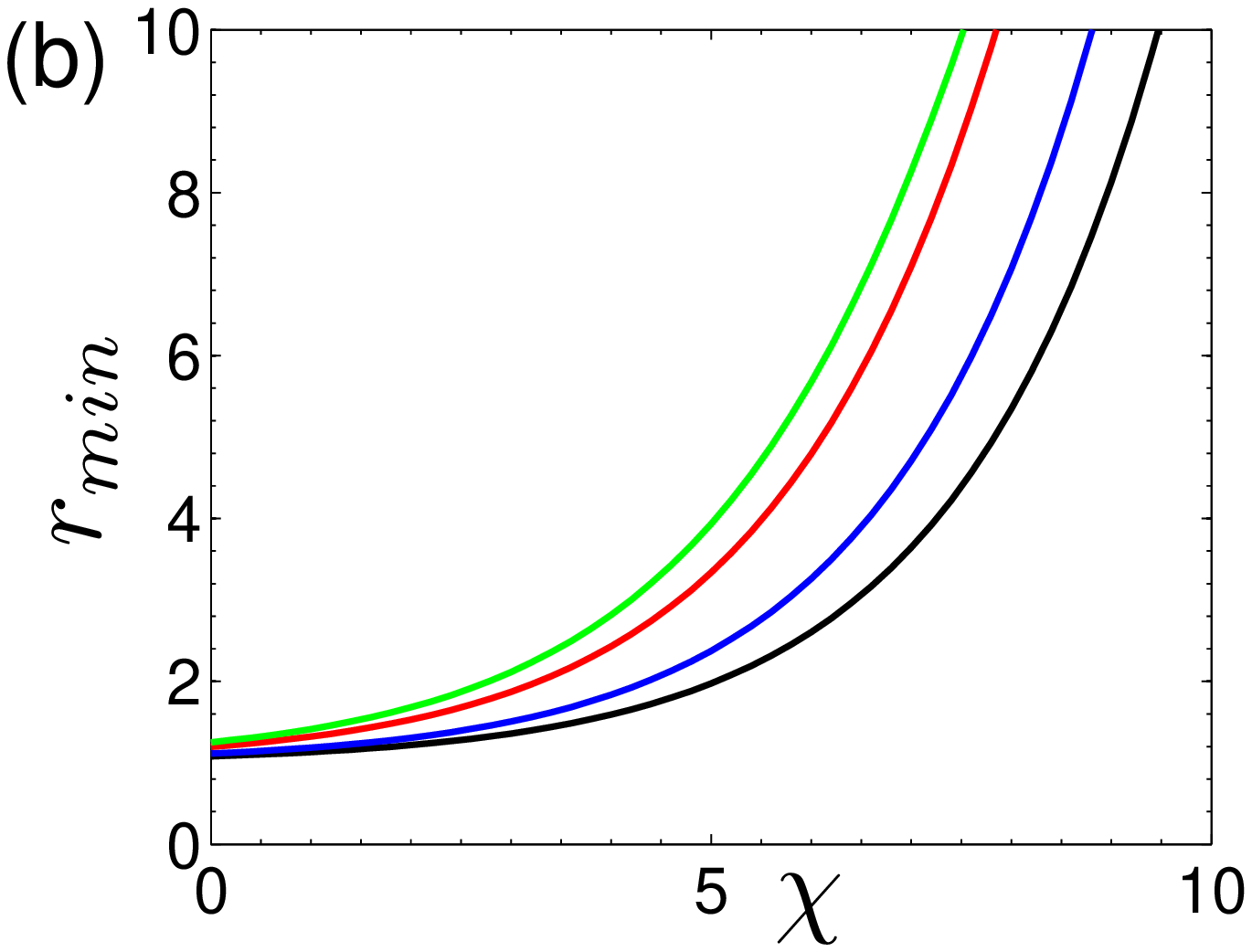}
      \includegraphics[scale=0.28]{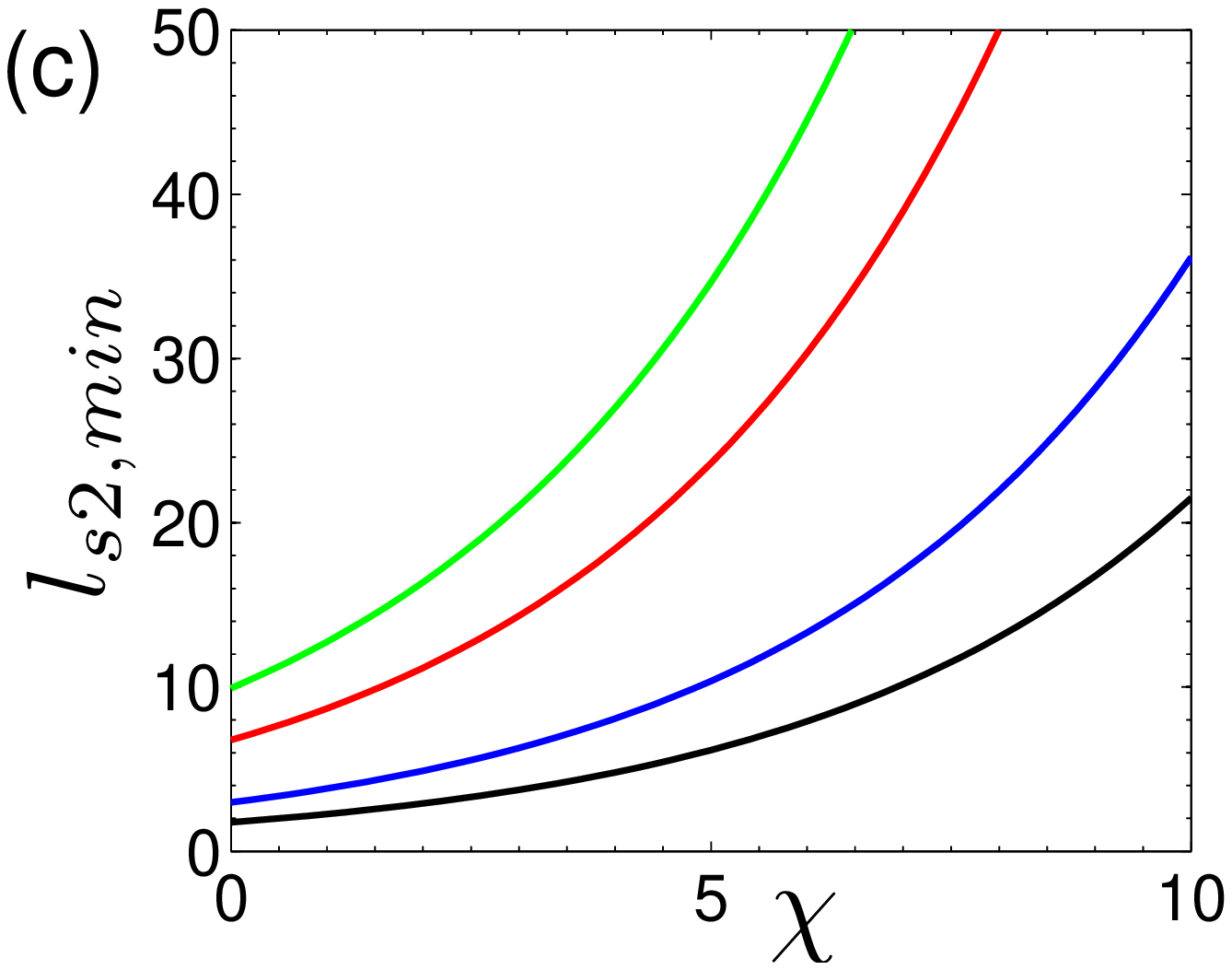}
       \includegraphics[scale=0.28]{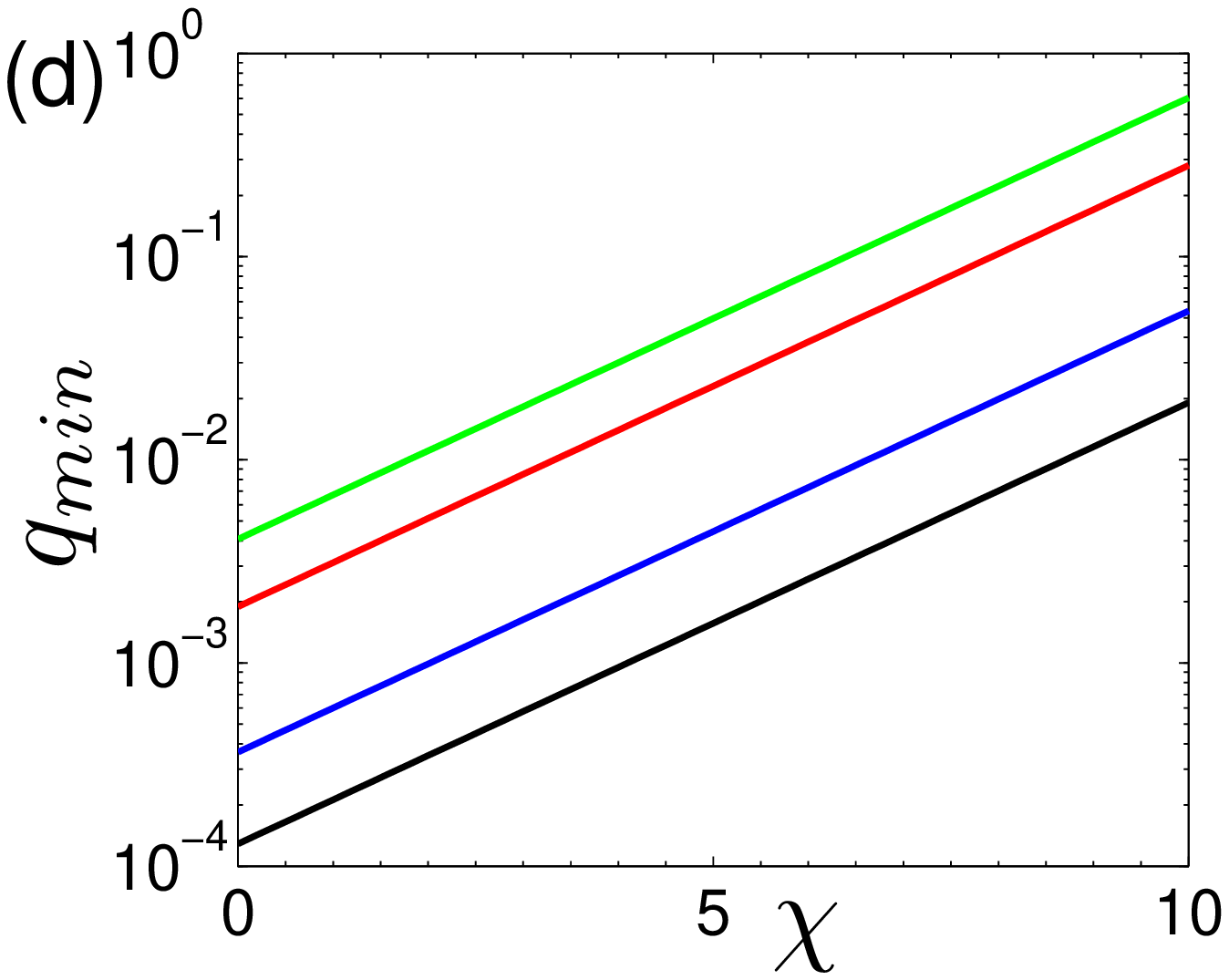}
       \caption{Optimal search process. Quantities are plotted as a function of $\chi$ and
       for various length $l_{s1}$. The parameter values are $L=2.4\times 10^6$bp, $k_{32}=(1.4
ms)^{-1}$, $D_2 =2 \mu m ^2s^{-1}$ and $D_1=D_2 e^{-\chi}$. The dashed curve in (a) corresponds to the minimal time with a single sliding state. $\tau_{min}$ is given in sec. }
\label{fig_tau_min}
\end{center}
\vspace{-0.8cm}
\end{figure}

We now proceed with some numerical estimations using parameters for
E.coli bacteria: $L=2.4 \times 10^6$bp (half the size of E. coli
DNA, to compensate that the target is located at the boundary in our
analysis), $k_{32}=(1.4\mu s)^{-1}$
\cite{Elf_Science2007,MalherbeHolcman_PLA2010} and $D_2=2
\mu m^2$, comparable to the 3D diffusion constant \cite{Elf_Science2007}. In Fig.~\ref{fig_tau_min}a, we plot the
minimum of $\tau$ as a function of $\chi$ and for various
$l_{s1}=(0.5,1,3,5)$ (in units of bp). The case $l_{s1}=0.5$ can be considered as an
effective description of a physical search process where a TF is
bound and immobile in state 1 (similar to the scenario considered in
\cite{HuGrosbergBruinsma_BJ2008}): after switching back to state 2,
the TF position in state 2 has changed only slightly in the range of a single
bp (the average of the maximum diffusion length in state 1 is $2
l_{s1}=2\sqrt{D_1/k_{12}}=1$). This position change can also be interpreted
as the variability due to the unbinding process. The mean
binding time $k_{12}^{-1}$  depends on the energy barrier $\Delta E$
(in units of $k_{B}T$) separating state 1 from 2. Comparing the
Arrhenius formula $k_{12} = \xi e^{-\Delta E} $, where $\xi$ is an
effective prefactor, with $k_{12} = D_1/l_{s1}^2=D_2
e^{-\chi}/l_{s1}^2$, we identify $\chi=\Delta E$ and $\xi=
D_2/l_{s1}^2$. Hence, for $l_{s1}$ small, the parameter $\chi$ is
the binding energy, however, for large $l_{s1}$, $\chi$ is related to the variance of the binding energy landscape in state 1,
as described in
\cite{ZwanzigPNAS1988,SlutskyMirny_BJ2004,MalherbeHolcman_PLA2010}.

Fig.~\ref{fig_tau_min}a shows that $\tau_{min}$ is initially not
very sensitive to $\chi$ until $\chi \sim
\ln\alpha$  (for $l_{s1}=0.5$ we have $\ln\alpha \sim 6$). In
contrast, with a single sliding state the minimum $\tilde
\tau_{min}= 2\sqrt{{L^2}/{(D_2 k_{32})} }e^{\chi/2}$ (with
$k_{31}=k_{32}$) increases exponentially with $\chi$ and quickly
reaches much higher values (black dashed curve in
Fig.~\ref{fig_tau_min}a). Furthermore, within the two states sliding
model, the novel feature is that the time ratio $r_{min}$ at the
minimum is not constant but increases with $\chi$
(Fig.~\ref{fig_tau_min}b). As a consequence, the experimental
findings that a TF spends more time bound to the DNA compared to
diffusing inside the nucleus \cite{Elf_Science2007} is now
compatible with an optimal search process. For example,  for
$l_{s1}=0.5$, the experimental results $\tau_{exp} \sim 350 s$ and
$r_{exp} \sim 5$ \cite{Elf_Science2007} are compatible with a value
$\chi\sim 8$ (Fig.~\ref{fig_tau_min}a-b). Because diffusion in state
1 slows down as $\chi$ rises, the sliding distance $l_{s2,min}$ and
the probability $q_{min}$ to switch from state 2 to state 3
increase, thereby reducing the probability of recurrently visiting
the same DNA site in state 1 (Fig.~\ref{fig_tau_min}c-d).
Surprisingly, a larger detaching probability $q_{min}$ does not lead
to a higher fraction of time spent in state 3, which is counter
intuitive ($r_{min}$ increases, Fig.~\ref{fig_tau_min}b).
\begin{figure}[h!]
\begin{center}
       \includegraphics[scale=0.28]{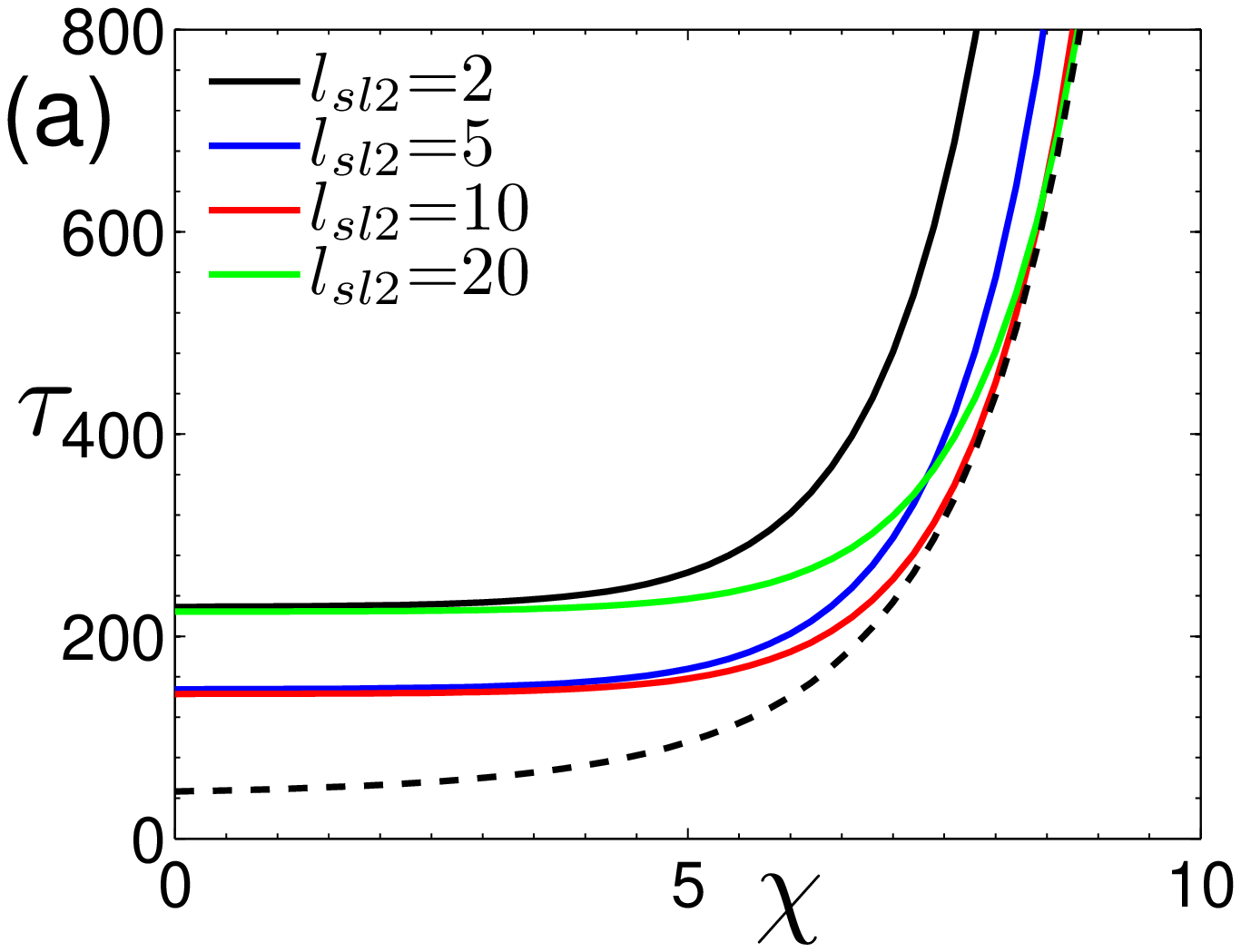}
       \includegraphics[scale=0.28]{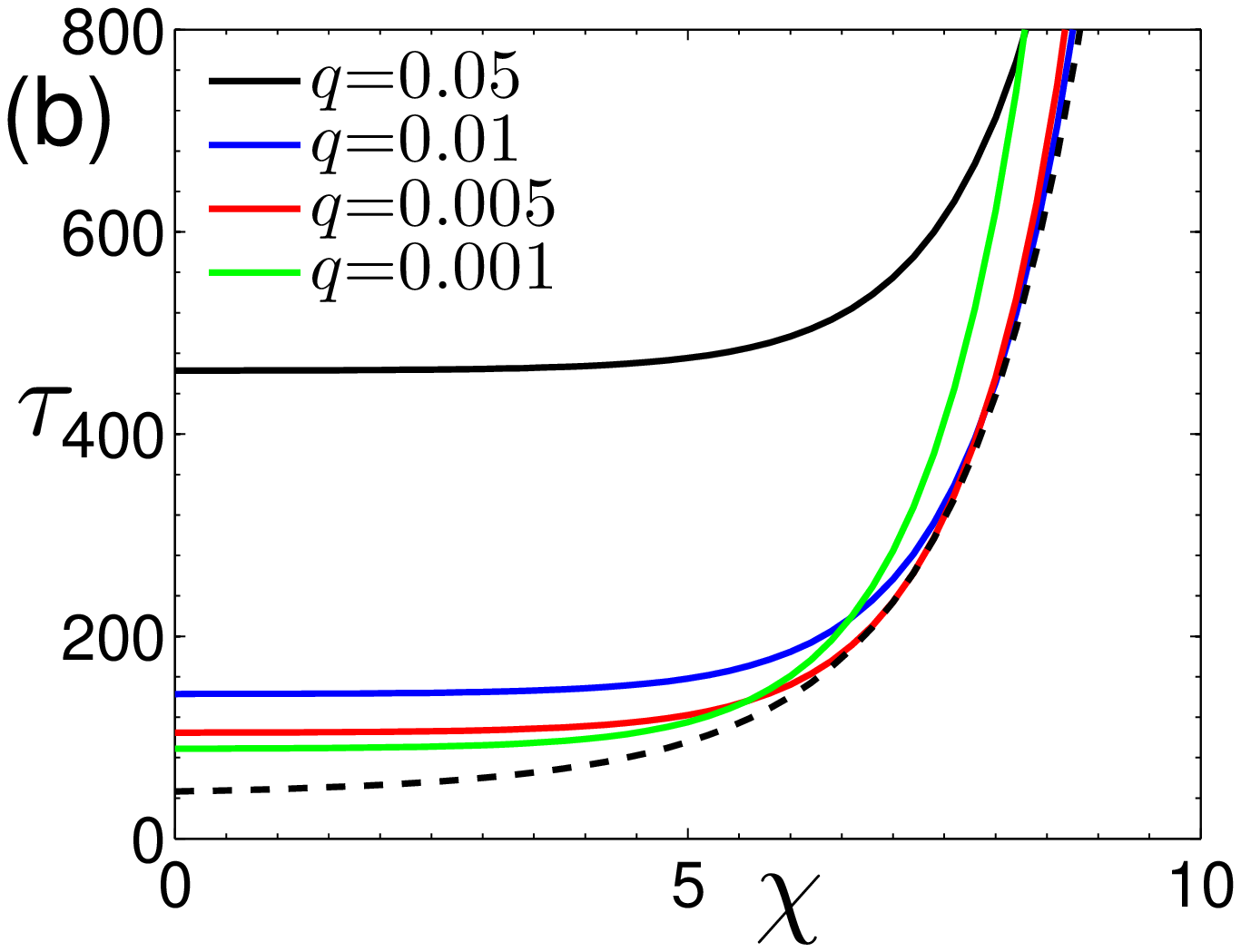}
       \includegraphics[scale=0.28]{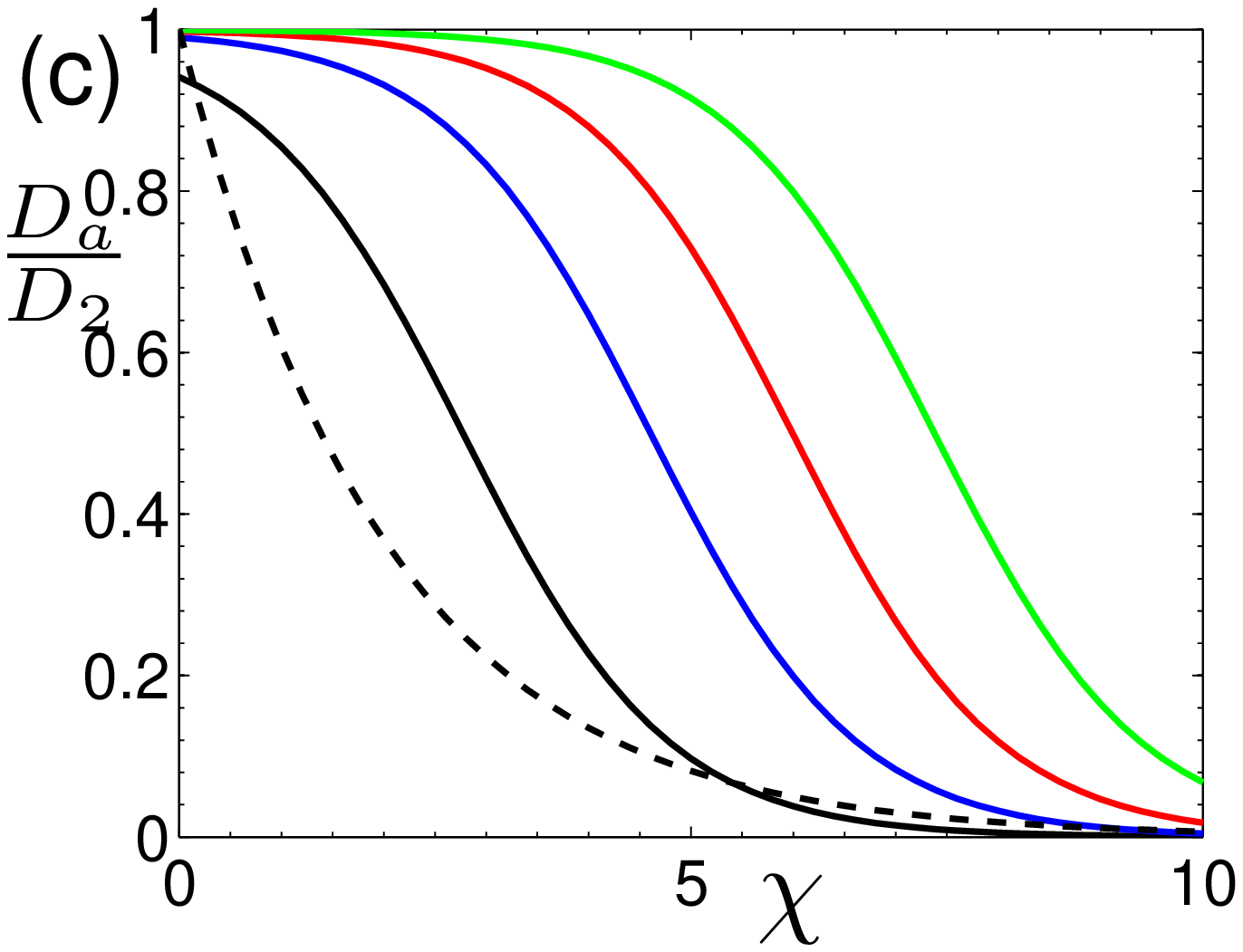}
        \includegraphics[scale=0.28]{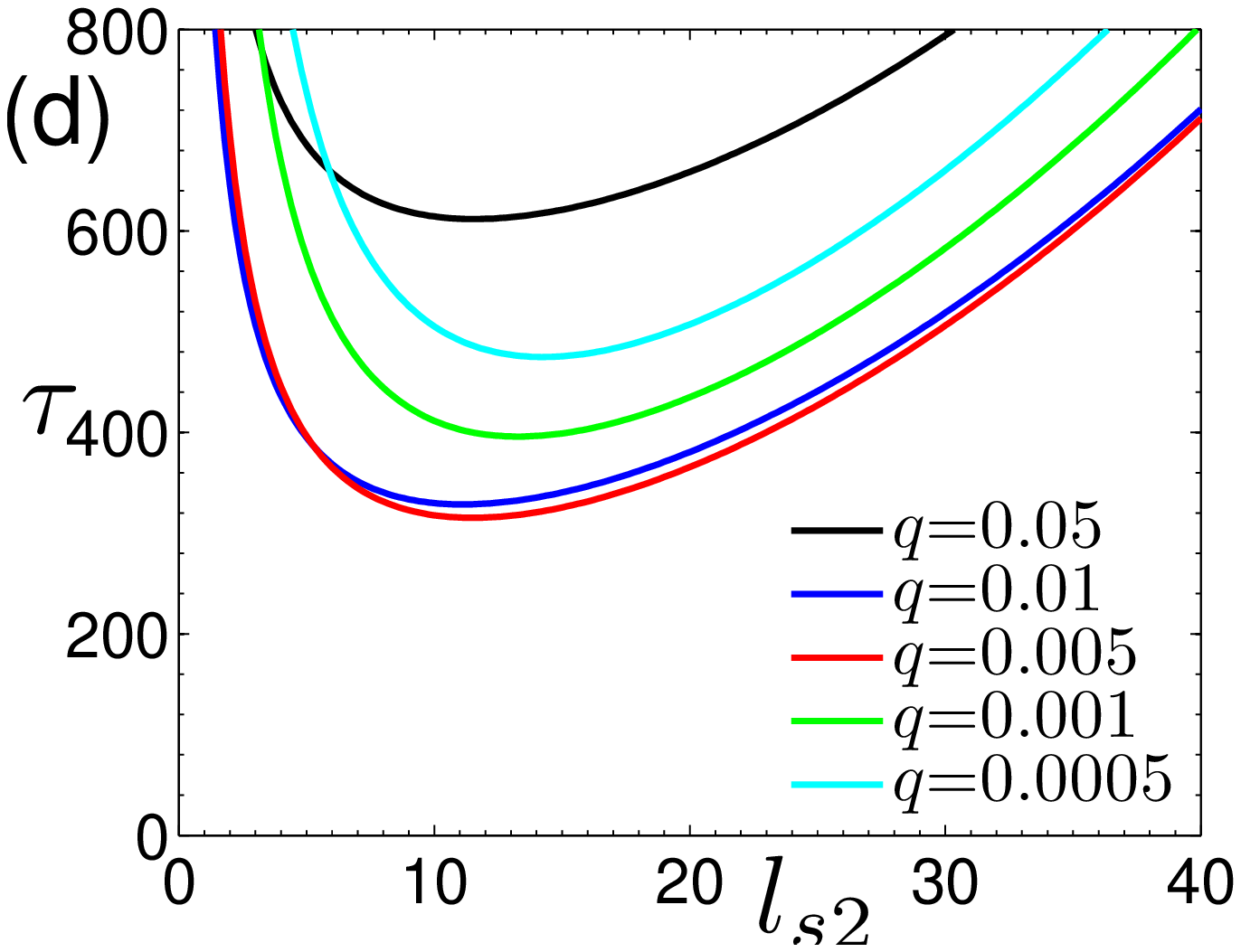}
       \caption{(a) Search time $\tau$ (in sec) for $l_{s1}=0.5$ and $q=0.01$. The
dashed line is the minimum of $\tau$. (b) $\tau$ for
$l_{s1}=0.5$ and $l_{s2}=10$. (c) Apparent diffusion constant
$D_{a}$ scaled by $D_2$ for the situation in (a). The dashed line is
the diffusion constant $D_1/D_2=e^{-\chi/2}$ for a model with a
single sliding state. (d) Search time $\tau$ for $\chi=7.5$ showing
a minimum. }
\label{fig_tau_1}
\end{center}
\vspace{-0.8cm}
\end{figure}
To study the impact of increasing the binding strength in state 1,
while the motion in state 2 (interaction with DNA backbone) is not
affected, we plotted $\tau$ as a function of $\chi$
(Fig.~\ref{fig_tau_1}a-b) for $l_{s1}=0.5$ and various $l_{s2}$ and
$q$ that are independent of $\chi$. This is in contrast with
Fig.~\ref{fig_tau_min}, where $\tau_{min}$ is achieved for values of
$q_{min}$ and $l_{s2,min}$ that do depend on $\chi$ and $l_{s1}$. In
Fig.~\ref{fig_tau_1}c, we plot the apparent diffusion constant $D_a$
(sliding along the DNA) as a function of $\chi$, with parameters
associated with panel a. $D_a$ decreases as $\chi$ increases, and
for $\chi \sim 7$, we have $D_a\sim 0.4\mu m^2s^{-1}$, which is
similar to measurements \cite{Elf_Science2007}. Within  a single
sliding state model, the 1D diffusion coefficient $D_1= D_2
e^{-\chi/2}$  decreases much faster as function of $\chi$ compared
to $D_a$ (dashed line in Fig.~\ref{fig_tau_1}c). We conclude that
experimental measurements of the apparent diffusion constant are compatible with much stronger binding energies in a two-state compared to a single state model. Finally, we show how $\tau$ is modulated by
varying $q$ or $l_{s2}$ for $\chi=7.5$ (Fig.~\ref{fig_tau_1}d).

To conclude, we showed here that the TF search time, characterized by switching between two states on the DNA, is considerably faster and less sensitive to binding energy fluctuations compared to a single 1D sliding state. Performing fast translocations ('hoppings') of the order of 10bp in state 2 speeds up the search time by reducing a slow recurrent search in state 1. In our analysis, switching between state 1 and 2 is a common and necessary feature of the search mechanism, in contrast to scenarios, where it is induced at strong DNA binding sites \cite{SlutskyMirny_BJ2004}.  State 2 further offers the possibility that a TF moves along the DNA by translation without the need to follow the double-helix rotation. Furthermore, since DNA promoter sequences are usually $\gtrsim 10$bps and even present in several copies \cite{WunderlichMirny_TrendsGenet2009,Laessig_GeneRegReview_2007}, small translocations in state 2 are unlikely to overshoot the target region. We show that an optimal search in our switching model involves a larger time spent bound to the DNA compared to diffusing in the nucleus, in agreement with experimental findings \cite{Elf_Science2007}. Finally, we find that the search time is very sensitive to changes in the detaching probability $q$. Hence, changing the TF interaction with the DNA backbone via modifying the electrical properties of the TF or the DNA by phosphorylation, methylation or acetylation is an efficient way to modulate the search time, and ultimately the cellular response. Future works should clarify the impact of the binding energy fluctuations in state 1, and should analyze in details the 3D dynamics, for example by considering DNA coiling \cite{LomholtBroeckMetzler_PNAS2009}. Moreover, in eukaryotes, the compact DNA structure \cite{Liebermanetal_FractalGenome_Science2009} and possible nuclear transport mechanism \cite{ShavtalGruenbaum_BiologRep2009} might as well be critical. Nevertheless, we expect that our results derived here remain a good approximation as long as subsequent attaching positions to the DNA are well separated compared to the average distance a TF slides along the DNA before detaching (around 100bp), and the time spent in 3D is approximately exponentially distributed, both of which are widely used and accepted in the literature.

{\bf Acknowledgement:} this research is supported by an ERC starting grant.

\bibliographystyle{apsrev}


\end{document}